\documentstyle[rmaaconf]{article}

\begin{document}

\title{Modern Numerical Hydrodynamics and the Evolution of Dense 
       Interstellar Medium}

\author{    Tomasz Plewa\altaffilmark{1}
        and Micha{\l} R\'o\.zyczka\altaffilmark{2}$^{,}$\altaffilmark{3}
       }

\altaffiltext{1}{Max-Planck-Institut f\"ur Astrophysik, Garching, Germany}
\altaffiltext{2}{Warsaw University Observatory, Warsaw, Poland}
\altaffiltext{3}{N. Copernicus Astronomical Centre, Warsaw, Poland}
\begin{resumen}
Se discuten los problemas relacionados con las simulaciones hidrodin\'amicas
en medios de alta densidad y se describe un nuevo c\'odigo num\'erico, basado
en la t\'ecnica de refinamiento de mallas, para simular la evoluci\'on de
remanentes de supernova en estos ambientes.
\end{resumen}
\begin{abstract}
We discuss some typical problems related to numerical hydrodynamics of
a dense interstellar medium.  A newly developed hydrodynamical code
based on adaptive mesh refinement technique is presented and applied
to simulate the evolution of a supernova remnant in a high-density
medium.  Advantages of this new approach in comparison to commonly
used hydrodynamical methods are briefly discussed.
\end{abstract}
%
%
%
%
%
%
\section{INTRODUCTION}
There is a growing evidence that some massive stars end their lives as
supernovae while being surrounded by a dense circumstellar medium
(Fransson 1994, Terlevich 1994). Most of the evidence comes from 
radio observations indicating that in some cases (SN~1979C,
SN~1980K, SN~1986J) the progenitors were probably losing their
envelopes, in dense stellar winds, shortly before the explosion 
(Van Dyk et al. 1996). In other cases (SN~1978K, SN~1988Z), a
strong radio remnant becomes a powerful X-ray source only a
few years after explosion (Ryder et al. 1993, Fabian \& Terlevich
1996). In both cases, the most probable origin of the emission is the
interaction of the supernova ejecta with an external medium.

To explain properties of the radio emission from supernova remnants (SNRs),
Chevalier (1982) proposed the {\em mini-shell\/} model in which
this emission originates from the Rayleigh-Taylor unstable shell
formed in the region separating the reverse and forward supernova shocks.
This model was successfully applied to the analysis of radio observations
of several SNRs (see Weiler et al. 1996 for a
summary). Hydrodynamical models of the ejecta-wind interaction phase
(Chevalier \& Blondin 1995) clearly show the rapid formation of a
dense shell in the shocked ejecta behind the reverse shock. In this
case the reverse shock becomes a source of the X-ray emission which
is partially reprocessed by unshocked ejecta and partially absorbed by
the shell (Fransson et al. 1996).
However, in its present form the ejecta-wind interaction scenario
fails to explain the very high (${\ga}10^{41}$ erg s$^{-1}$) X-ray
luminosity of SN~1988Z observed some 7 years after the explosion
(Fabian \& Terlevich 1996). There is also a compelling
spectroscopic evidence for the existence of a dense ($n_{\rm amb}\ga
10^6$ cm$^{-3}$ circumstellar medium (Stathakis \& Sadler
1991). Also, the slow decline rates of the broad-band
emission (Turatto et al. 1993) gives indirect evidence for an
additional source of the energy powering the remnant. As
it has been shown analytically (Shull 1980, Wheeler et al. 1980)
supernova remnants evolving in very dense ($n_{\rm amb}>10^{5}$
cm$^{-3}$) medium become powerful X-ray emitters shortly after the
blast wave begins to interact with the medium. This result has been
confirmed later by hydrodynamical simulations (Terlevich et al. 1992).
\section{NUMERICAL TECHNIQUE}

A major problem in numerical studies of supernova remnants is the large
range of spatial and temporal scales. For this reason it is difficult
to use conventional methods of astrophysical hydrodynamics to study such 
a violent phenomena in a dense medium. Here we present simulations of a 
SNR evolving in a uniform medium, performed with an adaptive mesh refinement
(AMR) technique.

The AMR can be regarded as an automatic procedure for the most efficient
discretization of the equations describing the evolution of the
physical system under study.  Here we deal with Euler equations
describing temporal and spatial evolution of the ejecta and the
interstellar medium. Since Euler equations are hyperbolic, they are
subject to the CFL condition setting the upper limit for the time
step, and the maximum allowed time step changes proportionally to the
size of the smallest zone of the grid.  This is a basic problem for
the algorithms that use moving meshes to increase the spatial
resolution. Moreover, due to the CFL limit it is practically
impossible to achieve proper resolution in two- and three- dimensional
simulations.
AMR builds a hierarchical (or nested) system of mesh patches located
at different levels. Each level forms a single grid which in turn may
contain one or more mesh patches. The patches may partially overlap or
may occupy separate regions of the computational volume, but every mesh
patch must be properly nested (covered) inside a patch located on the
parent grid. The grids have different effective resolutions; the ratio
of the zone size on the parent level to the zone size located on the child
level (refinement ratio) is an integer constant typically equal to 2
or 4. Since Euler equations are conservation laws (of mass, momentum,
and energy), the AMR algorithm must conserve physical quantities
advected between different patches located on the same grid as well as
between parent and child patches located on different grids.

Finally, the success of the AMR critically depends on the way in which
the new meshes are created. Different strategies have been proposed to
detect the regions which need to be refined in order to describe the
evolution of the system accurately. For example, one may try, using a
concept of Richardson extrapolation to estimate a local truncation
error of the scheme (Berger \& Colella 1988) but this method works
well only in regions free of shocks, since modern numerical schemes
always produce narrow shock profiles.  On the other hand, one may use
some empirical criteria if there is enough a priori information about the
physical behavior of the system. For example, to refine the shocks one
may try to detect the regions of compressive flows in which the local
pressure contrast exceeds some threshold value.
As the structure of the grids is dynamic, it consecutively adapts to
changing flow conditions, and the high resolution is used only in those
regions where the refinement criteria are satisfied. A full description of
the AMR algorithm is certainly beyond the scope of this
presentation. For present application we used the AMRA code (Plewa \&
M\"uller 1996) which closely follows a description of the AMR
algorithm as given by Berger \& Colella (1988); the hydrodynamical
equations are solved using the PPM algorithm by Colella \& Woodward
(1984).
\section{RESULTS AND DISCUSSION}
We simulate the expansion of the supernova ejecta into a homogeneous,
static medium in 1D assuming spherical symmetry. The mass of the
ejecta is equal to $M_{\rm ej}=5 M_{\odot}$, and the total energy of
the explosion is equal to $E_{\rm ej}=1\times10^{51}$ erg. For the
structure of the ejecta we adopt the profile given by Franco et
al. (1993). We assume that the medium has solar metallicity. Radiative
losses are calculated implicitly with a cooling function obtained with
CLOUDY 84.09 (Ferland 1993). The minimum temperature of
the gas is equal to $T_{\rm min}=10^4$ K.

At the symmetry axis ($R=0$), we impose a reflecting boundary condition
while at the outer edge of the grid ($r=1.79\times10^{17}$ cm) we
allow for a free outflow. The base grid (level=1) contains 512 zones
and we allow for two levels of refinement, with a refinement ratio
between the grids equal to 4. Thus, the maximum resolution (on
the third level) is equal to 8192 equidistantly distributed zones. The
meshes are created in the regions where the density or pressure contrast
exceeds 1.0 and 10.0, respectively. In other regions we control the
quality of the solution by using a truncation error estimate, with a
maximum allowed error of 0.01. In addition, we do not allow for a
refinement in the region occupied by the unshocked ejecta. We use a
Courant number equal to 0.8.

\begin{figure}
\vspace*{43mm}
\begin{minipage}{43mm}
\includegraphics{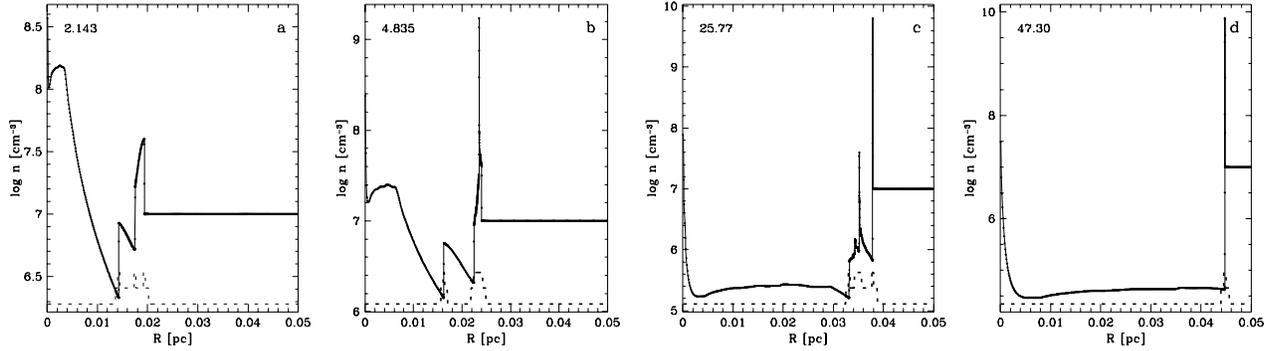}
\end{minipage}

\caption{
Temporal evolution of gas density and grid location (solid and dashed
lines, respectively) for a SNR evolving in a dense medium.
Evolutionary times (shown in upper-left corner of each panel) are given
in years.
}
\label{fig:1}
\end{figure}

The results of the simulation are presented in
Fig. \ref{fig:1}.\footnote{The movies presented during the talk can be
accessed at
{\tt http://www.MPA-Garching.MPG.DE/Hydro/AMR/movies/starbur.html}.}
Due to very high temperatures, the early expansion of the remnant
proceeds almost adiabatically, but after two years
(Fig. \ref{fig:1}a) the first signs of cooling are visible in the
post-main-shock region. At this time the finest grid covers the reverse
shock ($R\approx 0.0145$ pc), the contact discontinuity ($R\approx
0.0175$ pc), and the forward shock ($R\approx 0.0195$ pc). After next 2.7
years (Fig. \ref{fig:1}b), catastrophic cooling leads to the formation
of a dense shell right behind the main shock, and the whole region
between the contact discontinuity and the forward shock is covered by
the finest grid. Still later ($t\approx25$ yr) the shocked ejecta
starts to cool rapidly and the inner shell forms
(Fig. \ref{fig:1}c). Note that also in this case the region of strong
cooling is covered by the finest grid. Finally, both shells collapse
to form a single structure (Fig. \ref{fig:1}d).

We note that although the present resolution is much too low
to resolve some details (in particular both shells are unresolved), the
AMR algorithm is able to adequately follow the evolution of all
important flow structures. By calculating the mean volume occupied by
the finest grid during the whole evolution, we find that in this
particular case AMR offers an enormous saving of the CPU time (a
factor of 100). This makes us to believe that, at least for 
problems where the number of structures to be resolved occupy a
relatively small ($\le 10\%$) fraction of the domain, the adaptive
mesh refinement technique is an ideal tool. In the near future we plan
to extend to two dimensions using a more realistic
structure of the interstellar medium.

TP thanks the organizing committees for their hospitality during the
conference and for support from Programa Internacional de
Investigaci\'on en Astrof{\'{\i}}sica Avanzada {\em Guillermo
Haro}. TP also thanks Lynn for invaluable and warm contribution to his
work. The work of MR was supported by the grant KBN 2P-304-017-07 from
the Polish Committee for Scientific Research. The simulations were
performed on a workstation cluster at the Max-Planck-Institut f\"ur
Astrophysik.

\end{document}